\documentclass[prb,aps,twocolumn,superscriptaddress,floatfix]{revtex4-1}

\usepackage{color,graphicx,bm,amsmath}

\newcommand{\ld}{\ensuremath{L_{\mathrm{d}}}}
\newcommand{\jcf}{\ensuremath{\mathbf{j}_{\mathrm{CF}}}}
\newcommand{\Icf}{\ensuremath{I_{\mathrm{CF}}}}
\newcommand{\bnabla}{\ensuremath{\bm{\nabla}}}

\begin{document}

\title{Breakdown of counterflow superfluidity in a disordered quantum Hall bilayer}

\author{D. K. K. Lee}\affiliation{Blackett Laboratory, Imperial
College London, London SW7 2AZ, United Kingdom}

\author{P. R. Eastham}\affiliation{School of Physics, Trinity College, Dublin 2, Ireland.}

\author{N. R. Cooper}\affiliation{Cavendish Laboratory, 
University of Cambridge, Cambridge CB3 0HE, United Kingdom}

\date{September 1, 2010}

\begin{abstract} We present a theory for the regime of coherent
  interlayer tunneling in a disordered quantum Hall bilayer at total
  filling factor one, allowing for the effect of static vortices.  We
  find that the system consists of domains of polarized superfluid
  phase. Injected currents introduce phase slips between the polarized
  domains which are pinned by disorder. We present a model of
  saturated tunneling domains that predicts a critical current for the
  breakdown of coherent tunneling that is extensive in the system
  size. This theory is supported by numerical results from a
  disordered phase model in two dimensions.  We also discuss how our
  picture might be used to interpret experiments in the counterflow
  geometry and in two-terminal measurements.
\end{abstract}
\pacs{}

\maketitle

\section{Introduction}
\label{sec:intro}


In a quantum Hall bilayer at total Landau level filling $\nu_T=1$,
Coulomb interactions induce a state with interlayer phase
coherence~\cite{murphy_many-body_1994,lay_anomalous_1994}. This state
is expected to be approximately the Halperin [111]
state~\cite{halperin_mmm_1983}, which can be understood as a
Bose-Einstein condensate of interlayer
excitons~\cite{fertig_energy_1989,eisenstein_bose-einstein_2004}. The
motion of excitons corresponds to counterflowing electrical currents
in the layers, so that excitonic supercurrents can give
dissipationless electrical transport. The superfluid properties of the
[111] state have been demonstrated theoretically by Wen and
Zee~\cite{wenzee_neutral_1992,wen_superfluidity_2003}.

\begin{figure}[htb]
\includegraphics[width=0.4\textwidth]{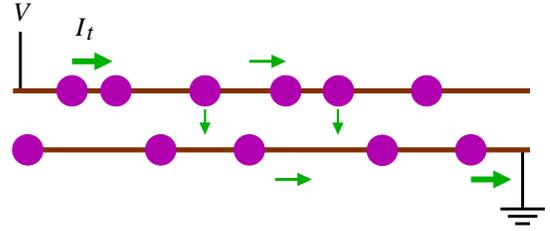}
\caption{Schematic diagram of tunneling experiment.
  \label{fig:schematic_tunnel}}
\end{figure}

This counterflow superfluidity can be probed in tunneling experiments.
In the tunneling geometry (Fig.~\ref{fig:schematic_tunnel}), a current
$I_t$ is injected into the top layer at one corner and removed from
the bottom layer at the opposite corner. These current flows may be
written as superpositions of layer-symmetric and layer-antisymmetric
currents,
\begin{eqnarray}
  I_{\mathrm{in(out)}}=\frac{1}{2}I_t\left[ \begin{pmatrix} 1 \\
      1 \end{pmatrix} \pm \begin{pmatrix} 1 \\
     -1 \end{pmatrix}\right],
\label{eq:currentcomponents}
\end{eqnarray}
where the two components refer to currents in the two layers. Thus,
the tunneling experiment corresponds to a flow of layer-symmetric
current, with equal counterflow currents $\Icf =I_t/2$ injected by
both the electron source and drain.  The symmetric component is
transported by a dissipationless edge state, which does not penetrate
the bulk due to an energy gap to charged excitations.  However, the
bulk can carry the counterflow component as a charge-neutral excitonic
supercurrent. Since both these channels are dissipationless, we expect
dissipationless electrical transport.  In particular, a finite
interlayer current $I$ at negligible interlayer voltage $V$ has been
predicted~\cite{wen_tunneling_1993,ezawa_lowest-landau-level_1993}.
This has been recently confirmed by
four-terminal measurements by Tiemann and
coworkers~\cite{tiemann_critical_2008,tiemann_dominant_2009}. This
phenomenon can be regarded as a form of the Josephson
effect~\cite{wen_superfluidity_2003}.
Note that thermally activated quasiparticles and contact
effects~\cite{su_critical_2010} can give rise to complications in
actual experiments.  

The Josephson-like regime persists for interlayer currents up to a
critical value $I_c$. Above $I_c$, interlayer transport becomes
dissipative.  Nevertheless, interlayer coherence can still be detected
in the interlayer $IV$ characteristics of the system.  A strong peak
is observed at zero bias in the differential interlayer conductivity.
This is followed at low bias by a regime with negative differential
conductivity~\cite{spielman_resonantly_2000,eisenstein_evidence_2003}. This
regime can be studied theoretically treating the interlayer tunneling
as a perturbation~\cite{jack_dissipation_2004,stern_theory_2001,balents_interlayer_2001}. 

In this paper, we focus on the Josephson regime below the critical
current, and present a physical picture of its breakdown. We have
previously presented, in a short paper~\cite{eastham_critcurr_2010}, a
theory of this breakdown based on numerical results on a
one-dimensional model. The aim of the present paper is to present
numerical results for a two-dimensional model, which directly
demonstrate the breakdown mechanism in a realistic geometry.  The key
motivation of our work is to understand the
observation~\cite{tiemann_dominant_2009} that the critical current
$I_c$ is proportional to the sample area.  (Area scaling is also
observed in the zero-bias peak of the interlayer
conductivity~\cite{finck_area_2008}. We will discuss this in
section~\ref{sec:discuss}.) The source and drain contacts for the
applied current are located at opposite ends of the system.  If one
models this system as a clean homogeneous bilayer using reasonable
estimates of the tunnel splitting, one finds that the injected current
should have tunneled across the bilayer within a few microns of the
source contact [$\lambda_J$ in Eq.~(\ref{eq:josephsonlength})]. Such a current
profile would suggest that the critical value of the interlayer
current should \emph{not} depend on the sample length in the direction
of the
current~\cite{fil_josephson_2009,abolfath_critical_2003,su_critical_2010}.
Put another way, the area scaling of the critical current could only
be explained by a clean model of the bilayer if one accepts a tunnel
splitting that is anomalously small by several orders of
magnitude~\cite{su_critical_2010}.

A similar puzzle is found in the original observation of dissipationless counterflow~\cite{tutuc_counterflow_2004,
kellogg_vanishing_2004} in
the counterflow geometry (Fig.~\ref{fig:schematic_leak}). Again, counterflow currents apparently traverse
the system over distances orders of magnitude further than expected. We will return to this geometry
in the final section. 

The resolution of this puzzle lies in the presence of disorder. We
shall see (Fig.~\ref{fig:2dmodel}) that, in the presence of static
phase disorder (pinned vortices), the supercurrent profile can be
pinned by disorder.  The time-independent supercurrents can then
penetrate into the sample over indefinitely large distances, limited
only by the finite size of the sample.  In fact, we find that
dissipation only appears when supercurrents completely fill the
sample.  This mechanism gives a critical current [Eq.
(\ref{eq:critcurrent})] which is proportional to the area of the
sample. The magnitude of this critical current agrees with
experiments, using reasonable estimates of the
parameters~\cite{eastham_critcurr_2010}.

This paper is organized as follows.  We will discuss the origin of
disorder in the bilayer in Section \ref{sec:disorder}.  Then, in
section \ref{sec:phasetheory}, we will introduce the phase Hamiltonian
for the excitonic superfluid that describes the interlayer-coherent
phase of the quantum Hall bilayer.  In section \ref{sec:pinning}, we
discuss how quenched vortices in the superfluid affect the ground
state of the system and its response to injected currents. Then, in section \ref{sec:numerics}, we
present the results of a numerical simulation of the bilayer in the tunneling configuration to support the prediction of our theory.
In the final section, we discuss how our picture can be used to interpret experiments for the bilayer in other configurations.

\section{Model of Disorder}
\label{sec:disorder}

Weak disorder, such as a spatially varying tunneling splitting, does
not affect the tunneling properties of the system
dramatically~\cite{roslee_2010}. A tunneling mechanism based on a
disordered edge has been proposed by Rossi \emph{et
al}~\cite{rossi_interlayer_2005}. However, such a theory predicted
linear scaling with the sample length but not its area.

We consider here a bilayer with charge disorder in the bulk.  One
common source for this disorder is the electrostatic potential due to
disordered dopant layers.  We expect the incompressible quantum Hall
phase to occupy only a fraction of the sample, with the remainder
occupied by puddles of compressible electron liquid. Thus, the
incompressible phase forms a network of channels separating puddles of
size $\xi \approx d_d \approx\mathrm{200\ nm}$, the distance to the
dopants. We suppose that the width of the channels is of the order of
the magnetic length $\ell_B\sim 20$ nm.  This coherent network model
was first studied in the context of the quantum Hall bilayer by Fertig
and Murthy~\cite{fertig_coherence_2005}.

In a quantum Hall superfluid, excess charge nucleates vortices in the
exciton
superfluid~\cite{eastham_vortex_2009,fertig_coherence_2005,stern_theory_2001,balents_interlayer_2001,fogler_josephson_2001}.
For a balanced bilayer with individual layer fillings
$\nu_1=\nu_2=1/2$ these vortices are merons of charge $\pm e/2$. (In an
unbalanced bilayer the charges are~\cite{roostaei_theory_2008} $\pm
e\nu_{1(2)}$.) In previous work~\cite{eastham_vortex_2009}, we have
studied how the vortex density is determined by a competition between
the superfluid energy cost of the vortex and the charging energy of
each puddle. We found that the bilayer can be strongly disordered in
the current experimental regimes. This suggests that the random field
due to the pinned vortices has an exponentially decaying correlation
function in space.

The above scenario provides a specific physical model for quenched
vortices with short-ranged correlations in the exciton superfluid.
The theory we present below depends on the existence of trapped
fractional $e/2$ charges to create these vortices but does not depend
crucially on the details of the disorder distribution. Our results
should be valid as long as the vortices are dense enough that their
separation ($\sim \xi$) is smaller than the clean tunneling length scale
$\lambda_J$ [Eq.~(\ref{eq:josephsonlength})].

\section{Phase model}
\label{sec:phasetheory}

In the previous section, we have outlined a model of disorder which induced quenched vortices in a quantum Hall state.
To describe this exciton superfluid with quenched vortices, we start with an effective Hamiltonian
for the phase $\theta$ of the superfluid. We separate out the component, $\theta^0$, of the phase field that is due to 
the quenched vortices. The remaining component, $\phi\equiv \theta-\theta^0$, would have no vorticity in the ground state but may acquire vorticity in the presence of injected currents and other external perturbations.
It can be shown that the effective Hamiltonian can be written as a random field XY model:
\begin{equation}
H_{\rm eff}
  =\int \left[\frac{\rho_s}{2} (\bnabla \phi)^2 - t \cos
    (\phi+\theta^0) \right]d^{D}\mathbf{r},
  \label{eq:phaseenergy_phi}
\end{equation} 
which describes the low-energy phase fluctuations of a bilayer
containing pinned vortices. This form is a simple extension of the
form for a clean model~\cite{wen_tunneling_1993}.  The first term
describes the superfluid stiffness to phase twists while the second
describes the interlayer tunneling.  We will assume that the quenched
phase field $\theta^0$ has a correlation length of $\xi$.

In the Josephson regime, there is no quasiparticle flow at zero
temperature. All currents are accounted for by superflow and coherent
tunneling. The counterflow supercurrent density above the ground
state, $\jcf$, and the interlayer tunneling current density, $J_t$,
are related to the phase field by:
\begin{equation}
\jcf=\frac{e\rho_s}{\hbar} \bnabla \phi\,,\qquad
J_t=\frac{et}{\hbar}\sin (\phi+\theta^0)\,.
\label{eq:currents}
\end{equation}

A time-varying superfluid phase $\phi(t)$ gives rise to an interlayer
voltage difference $V$ \emph{via} the Josephson relation
\begin{equation}
V=\frac{\hbar\dot{\phi}}{e}.
\label{eq:josephson}
\end{equation}
  Therefore, a state with a finite interlayer
current at zero interlayer voltage is time-independent, 
corresponding to a local minimum of the energy (\ref{eq:phaseenergy_phi}).
The stationary equation
is simply the continuity equation stating
that the loss of counterflow current is accounted for by
interlayer tunneling: $\bnabla\cdot \jcf = J_t$.
This can be written as
\begin{equation}
-\rho_s\nabla^2\phi + t \sin(\phi+\theta^0) = 0\,.
\label{eq:static}
\end{equation}
All states with zero interlayer voltage obey this equation. The
dependence on the injected current arises as the boundary conditions
at the source and drain specifying the injected counterflow component
$\jcf$.  In terms of the phase field, this is a boundary condition on
$\bnabla\phi$.

We expect that the counterflow current injected at the boundary will
decay into the sample because interlayer tunneling will recombine
electrons and holes across the two layers, as depicted in
Fig.~\ref{fig:schematic_leak}.  In the clean case ($\theta^0=0$), one
expects~\cite{fil_josephson_2009,bak_commensurate_1982} the static
solution to show all the injected counterflow current tunneling across
the bilayer over a ``Josephson length'' of
\begin{equation}
\lambda_J = \sqrt{\rho_s/t}\,.
\label{eq:josephsonlength}
\end{equation}
This length scale is estimated to be of the order of a few microns using realistic parameters.

Since the phase angle is compact, this implies a maximum injected
current density of $\rho_s|\bnabla\phi|\sim \pi\rho_s/\lambda_J$. For higher injected currents,
phase slips enter and propagate through the system. This gives rise to a time-varying
phase and hence a non-zero interlayer voltage \emph{via} the Josephson relation 
(\ref{eq:josephson}).

Note that this picture of current penetration into the clean system
gives a penetration depth as a microscopic length scale independent of
the injected current. We will see below that the disordered system
behaves qualitatively differently --- the current can penetrate into an
indefinitely large area of the system. The reason is that injected
phase slips are pinned by disorder and therefore a static solution to
Eq.~(\ref{eq:static}) can persist to higher injected currents.
In the next section, we will discuss this picture of pinning.

\section{Pinned superfluid}
\label{sec:pinning}

We will now review the heuristic theory of pinning presented in our
previous work~\cite{eastham_critcurr_2010} in order to provide the
context to interpret our simulation results.  The quenched vortices
play a crucial role for the critical current.  They pin any injected
supercurrents and sustain dissipationless states. This is reminiscent
of how disorder pins magnetic flux in
superconductors~\cite{tinkham_introduction_1996,larkin_pinning_1979,
  vinokur_collective_1990}, or charge in charge-density
waves~\cite{fukuyama_dynamics_1978}. However, we emphasize that there
is a significant difference in the bilayer compared to these other systems.
In the superconductor, the depinning force arises from the Lorentz
force on the flux lines due to the bulk transport current. In
charge-density-wave systems, depinning originates from the electric
field in the bulk which is an insulator when the charge density cannot
slide.  In the quantum Hall bilayer, depinning arises from the
injected charge current which is applied \emph{only} at the sample
boundary. Thus, in this case, the critical current will depend on how
the depinning ``forces'' are transmitted through the system.  In such
a geometry, it is not immediately obvious how the critical current
$I_c$ would scale with the area of the whole sample.

We will borrow from the Fukuyama-Lee theory~\cite{fukuyama_dynamics_1978} of disordered charge density waves and the Imry-Ma theory~\cite{imry_random-field_1975} for ferromagnets in random fields. We recall the
form of the ground states of the random field XY model, Eq.~(\ref{eq:phaseenergy_phi}), in the weak disorder regime
$\xi\ll\lambda_J$ relevant for the bilayer. In this regime, it is energetically costly for the phase $\theta$
to follow the random field $\theta^0$ which varies over the scale of the correlation length $\xi$.
The ground state consists of domains of polarized phase. These domains cannot be arbitrarily 
large because the energy cost of the mismatch between the phase and random field grows with the domain size.
The energy cost for a
phase twist that varies over the scale $l$ is $E_s(l) \sim \rho_s
l^{D-2}$ in $D$ dimensions. The typical tunneling energy of a polarized region of size $l$ is
obtained by summing random energies in the range $\pm t\xi^D$ for its
$(l/\xi)^D$ correlation areas, giving $E_t(l) \sim t\xi^D(l/\xi)^{D/2}$. 
The phases will twist when $E_t(l)$ exceeds $E_s(l)$. Therefore, the ground state consists of domains of size $\ld$ determined by
\begin{equation}
E_s(\ld) \sim E_t(\ld).
\label{eq:energybalance} 
\end{equation}
This ``Imry-Ma scale'' for the domain is:
\begin{equation} 
\ld\sim
\left(\frac{\rho_s}{t\xi^{D/2}}\right)^\frac{2}{4-D}=\left(\frac{\lambda_J^2}{\xi^{D/2}}\right)^{\frac{2}{4-D}}.
\label{eq:domainsize}
\end{equation} 
In this ground state of polarized domains, the average coarse-grained phase over a domain is chosen such that the tunneling energy $H_t$ of each domain is minimized.
Since $\delta H_t/\delta\phi(\mathbf{r})$ is the tunneling current at position
$\mathbf{r}$, the total tunneling current over the domain vanishes. 

In two dimensions, $\ld = \lambda_J^2/\xi$. We see that, in the
experimentally relevant regime of $\lambda_J \gg \xi$, this new
disorder-induced length scale is much larger than the Josephson
length. This is the length scale controlling current penetration into
the sample. However, as we see below, this should not be interpreted
simply as a renormalized length scale for how far counterflow currents
penetrate into the sample.

Consider now the effect of an injected counterflow which imposes a
phase twist at the boundary.  The phase will therefore twist away from
its equilibrium configuration.  We assume that the domain at the boundary
remains polarized at short distances and so will rotate uniformly on
the scale of $\ld$.  This generates a tunneling current which reduces
the counterflow current.  The residual counterflow currents will be
transmitted further into the sample, causing the domains there to
rotate in a similar way.

This picture allows us to average over each domain.  The total 
tunneling current in a domain consists of a similar random sum to that for the tunneling energy, $E_t$, 
and is given by $I_{\rm d} f(\bar\phi)$ where
\begin{equation}
I_{\rm d} = \frac{eE_t(\ld)}{\hbar} = \frac{e\rho_s}{\hbar} \ld^{D-2},
\label{eq:domaincurrent}
\end{equation}
$\bar\phi$ is the
deviation of the coarse-grained phase from its equilibrium value, and
the range of $f(\bar\phi)$ is typically $[-1,1]$. 
To minimize the region pushed out of equilibrium by the injected current, each domain will rotate so as to minimize the counterflow current transmitted into the sample. This maximizes the tunneling current and is achieved if we choose $|f|\sim 1$. Thus, we argue that forcing at a boundary
leads to a self-organized critical state, in which the driven part of
the system is saturated at the threshold $|f|\sim 1$. This means that the area $S_t$ 
of the system driven out of equilibrium to provide coherent tunneling is simply proportional to the number of domains
necessary to carry the injected current $I$. Each domain can support a current of $I_{\rm d}$ and so
\begin{equation} 
\frac{S_t(I)}{\ld^D} \approx \frac{I}{I_{\rm d}}.
\label{eq:areacurrent} 
\end{equation}
The critical current is reached when all domains in the sample are saturated: $S = S_t(I_c)$ for a 2D sample of area $S$.
Therefore the critical current for a sample of area $S$ is
\begin{equation} 
I_c\sim I_{\rm d} \frac{S}{\ld^D} = \frac{e \rho_s}{\hbar}  \frac{S}{\ld^2}, 
\label{eq:critcurrent} 
\end{equation}
This formula also applies to the 1D case with $S$ being the sample length. 

\section{Numerical results}
\label{sec:numerics}

We will now present numerical results to support the theory in the previous section.
Our numerical results are obtained using the dissipative
model 
\begin{equation}
 -\lambda \dot\phi=\frac{\delta
    H_{\mathrm{eff}}}{\delta \phi} = -\rho_s \nabla^2\phi+t \sin
  (\phi+\theta^0) ,
\label{eq:bilayerphased}
\end{equation}
whose stationary solutions $\dot \phi=V=0$ are the local minima
of Eq.~(\ref{eq:phaseenergy_phi}). This is performed on a lattice model. 
The phase field $\theta^0_i$ at site $i$ is uncorrelated with the phase field at any other site.
This corresponds to taking the lattice spacing to be the correlation length $\xi$ of the original continuum model.
The natural unit of current is $I_0=e\rho_s/\hbar$.  
The results that we present below are the results for a 200$\times$20 lattice, averaged over 500 realizations of the disorder.
For this illustration, we take the ratio of the tunneling strength to the superfluid stiffness to be $t\xi^2/\rho_s=0.6$.
This corresponds to a Josephson depth $\lambda_J$ of the order of a lattice spacing and a domain size $\ld$ of 2 lattice spacings.
Although this is not deep in the weak-disorder regime considered in the previous section, 
our results appear to support the conclusions in that section.  

\begin{figure}[hbt]
\input{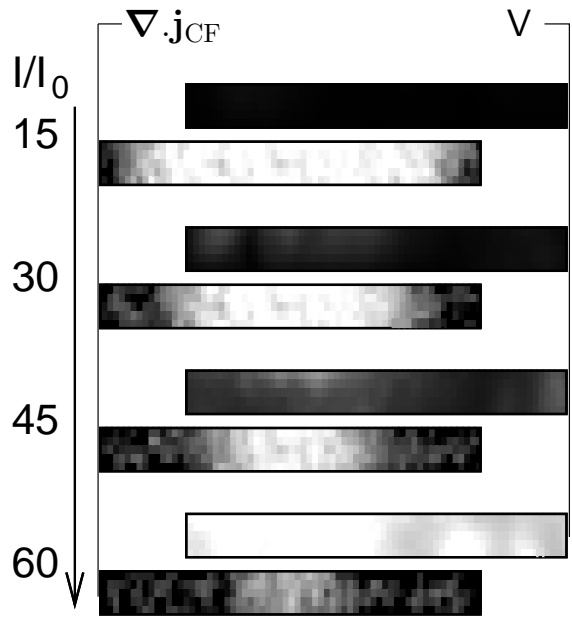}
\caption{Spatial distribution of tunneling currents (left column) and
interlayer voltages (right column), in a lattice model of
200$\times$20 sites, with current injection at the two lower
corners. The injected counterflow currents are $I/I_0 = 15, 30, 45,
60$ for the four pairs of plots. Dark colours, left column: high
current. Dark colours, right column: low voltage. $t\xi^2/\rho_s =
0.6$. Results are averaged over 500 realizations; tunneling currents
are summed over blocks of 3x3 sites.
\label{fig:2dmodel}
}
\end{figure}

The boundary conditions for Eq. (\ref{eq:phaseenergy_phi}) are
determined by the current flows through the
sample~\cite{su_critical_2010}.  We consider a tunneling geometry in
which, as seen in Fig.~\ref{fig:2dmodel}, a current $I_t$ is injected
into the top layer at the bottom left corner and removed from the
bottom layer at the bottom right corner. As already discussed
[Eq. (\ref{eq:currentcomponents})], the counterflow component of the
currents corresponds to equal counterflow currents $\Icf$ injected by
\emph{both} the electron source and drain.

The ground state of the system is found by evolving from a random state using the dissipative dynamics (\ref{eq:bilayerphased})
with the boundary condition of no injected current. From Eq.~(\ref{eq:currents}), this corresponds to 
$\hat{\bf n}\cdot\bnabla\phi=0$ everywhere on the boundary with $\hat{\bf n}$ being the normal to the boundary.
To model the current injection in a
tunneling experiment, we then slowly increase the counterflow current at the source and drain sites (1 and 2)
to the final values
$\xi\hat{\bf n}\cdot\bnabla\phi |_1=\xi\hat{\bf n}\cdot\bnabla\phi |_2=I/I_0$.
For the low values of the injected current $I$, the dynamics reach a static
solution, corresponding to the Josephson regime with vanishing
interlayer voltages. At higher currents, these time-independent
solutions break down and the phase winds continuously in time. 
This corresponds to the breakdown of the d.c.~Josephson regime
and the appearance of a state with finite interlayer voltages.

We expect that the counterflow current injected at the boundary will
decay into the sample because interlayer tunneling will recombine
electrons and holes across the two layers.  We find that the manner in
which this occurs is qualitatively different in clean and disordered
bilayers.  As mentioned in section \ref{sec:intro}, the penetration depth of the injected
current is simply the Josephson length $\lambda_J$ in the clean
case. We see in Fig.~\ref{fig:2dmodel} that, for the disordered case,
current penetrates further and further into the sample as we increase
the injected current from the two ends.  We see from the border of the
regions with finite tunneling ($\bnabla\cdot\jcf\neq 0$) that the
counterflow region increases linearly in area ($S_t$) with the
injected current. This is consistent with the prediction
[Eq. (\ref{eq:areacurrent})] for $S_t$ as a function of the injected
current from our theory.

At a high enough injected current ($I/I_0\simeq 50$), the current profiles 
from the contacts (lower left and right corners) will meet in the middle of the lattice.
Beyond this point, further increases in
current cannot be accommodated by coherent tunneling and an interlayer
voltage develops. 

We emphasize that this interpretation of the threshold for the
breakdown of the stationary solutions is qualitatively different from
the clean case.  In the clean model, the breakdown can be understood
in terms of the injection of phase solitons at the
boundary~\cite{littlewood_metastability_1982,fil_josephson_2009} when
the injected current exceeds the superflow that can be supported by a
static phase twist $|\bnabla\phi| \sim \pi/\lambda_J$.  These phase solitons
propagate through the sample. Thus, the phase at any point varies in
time, and the system develops an interlayer voltage by the
a.c.~Josephson effect. In this language, we can say that these
injected solitons can be pinned by disorder so that stationary
solutions exist even when there are many solitons in the system.

\section{Discussion}
\label{sec:discuss}

\begin{figure}[hbt]
\includegraphics[width=0.4\textwidth]{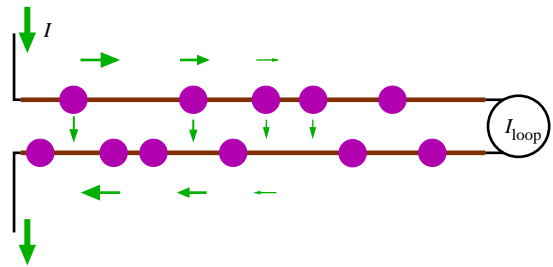}
\caption{Schematic diagram of a counterflow experiment with a short
circuit to complete current loop for counterflow. $I_{\rm loop}$
measures current through the short circuit. Diagram depicts the
Josephson regime where the loss of counterflow current through
tunneling means that $I_{\rm loop}=0$. The current-carrying region
penetrates to the right as the injected current $I$ is increased,
eventually reaching the other end at $I_c$.  }
\label{fig:schematic_leak}
\end{figure}

We have so far focused our discussion on the bilayer in the tunneling
geometry.  Finally, we will discuss how two other experimental
situations can be interpreted in our theory.  The first setup is the
transport in the bilayer in a counterflow geometry where the source
and drain contacts are on the same side of the bilayer while the other
end is short-circuited to allow the current to flow from the top layer
to bottom without the need for tunneling.  This is depicted
schematically in Fig.~\ref{fig:schematic_leak}.  This was first
investigated by Tutuc \emph{et al}~\cite{tutuc_counterflow_2004} and
Kellogg \emph{et al}~\cite{kellogg_vanishing_2004}.  A large current
($I_{\rm loop}$ in figure) was found passing through the short circuit
that join the top and bottom layers.  This seems to imply that there
is no leakage by tunneling across hundreds of microns.  As we
discussed in section \ref{sec:intro} for the case of the tunneling
geometry, a realistic estimate of the tunneling rate based on a clean
bilayer predicts that the injected current would have tunneled across
the bilayer within a micron and \emph{no} current should remain at the
far end.

In our theory, this situation can be simulated by solving
Eq. (\ref{eq:bilayerphased}) with injected counterflow current at one
end only, say the left end of Fig.~\ref{fig:2dmodel}. We expect the
tunneling domains to saturate successively from this end, and the
current profile is the same as that found in Fig.~\ref{fig:2dmodel}
for this side of the sample. There will be no current flow on the
right side. In other words, we expect to see zero current in the
short-circuit loop ($I_{\rm loop}=0$) in the Josephson regime. As we increase the
injected current $I$ to $I_c$, the current-carrying region reaches the
other end of the sample. Any further currents will pass through the
short circuit. For an ideal loop, we expect $I_{\rm loop} = I -
I_c$. 
However, the short circuit itself should have a finite
resistance. Therefore, the presence of a non-zero $I_{\rm loop}$
implies a small interlayer voltage at the end of the sample. In the
phase theory, the Josephson relation (\ref{eq:josephson}) means that
the superfluid phase must wind in time. Thus, a static solution to
Eq. (\ref{eq:bilayerphased}) becomes impossible anywhere in the system
and the whole sample develops an interlayer voltage. We expect that
the phase dynamics will be complex and chaotic. The nature of the
steady state would depend on details of the damping mechanisms. This
provides a zero-temperature picture of the counterflow geometry and is
consistent with the recent experiments of Yoon \emph{et
  al}~\cite{yoon_interlayer_2010}, in which the loop current $I_{\rm
  loop}$ is negligible for tunnel currents below a critical value. We
should keep in mind that, at finite temperatures, there may be
in-plane resistances associated with the flow of thermally activated
quasiparticles. 

The second situation we wish to discuss is the two-terminal
measurements of zero-bias interlayer conductance, $G(0)$, by Finck
\emph{et al.}~\cite{finck_area_2008}  Within our theory, this could be
interpreted as the dissipative regime (\emph{i.e.} no static solutions
for the phase) above the critical current. We note that these
measurements were performed close to the phase boundary between the
excitonic superfluid with interlayer coherence and the incoherent
$\nu=1/2$ quantum Hall liquid. We expect the critical current to
vanish near the phase
boundary~\cite{tiemann_dominant_2009,eastham_critcurr_2010}. Therefore, it is easy to exceed
the critical current in this region of the phase diagram. Then, we see
in Fig.~\ref{fig:2dmodel} that a non-zero interlayer voltage develops
across the whole sample and the tunneling current exists over the
whole sample. In other words, $G(0)$ should be proportional to the
area of the sample, as seen by Finck \emph{et al}.  We point out that,
whereas this interpretation gives an intrinsic zero-temperature source
of a finite conductance, there may be other sources of
dissipation. Even below $I_c$ there could be a finite dissipation due
to contact resistances and thermal activated vortex motion.
Fluctuations in the pinning energies could also lead to very weakly
pinned regions in large samples\cite{coppersmith_phase_1990}, which
may lead to dissipation below $I_c$ even at $T=0$.

In summary, we have presented a theory of the Josephson regime of
coherent tunneling in a disordered quantum Hall bilayer with static
pinned vortices. We find that, in the tunneling geometry, there are
two current-carrying regions emanating separately from the source and
drain contacts.  In these regions, coherent tunneling is
saturated. All injected counterflow current is lost by tunneling at
the edge of these regions.  The area of the saturated region $S_t$
grows linearly with the injected current $I$.  This linear relation is
predicted by our theory and is confirmed by the numerical results
presented here.  This is analogous to the Bean critical state for flux
penetration into a disordered superconductor.

This picture tells us that the system reaches the critical current
when the whole sample is saturated with coherent tunneling.  This
results in a critical current that is extensive for sufficiently large
samples that contain many domains of polarized phase.  In contrast,
the clean limit~\cite{su_critical_2010} sees area scaling for $I_c$
only for small samples (small compared to the Josephson length).

Theoretically, our results are qualitatively different from clean
theories~\cite{su_critical_2010} because of the existence of these
pinned polarized domains.  The size $\ld$ of these domains is a
disordered-induced length scale that emerges in our theory
[Eq.~(\ref{eq:domainsize})]. This scale has no counterpart in the
clean system. It would be therefore be useful if this length scale can
be probed in experiments.  We note that, for the area-scaling formula
(\ref{eq:critcurrent}) to apply, the sample should be large enough
to include many complete domains. 
For sample dimensions smaller than $\ld$, the system should cross over
to a regime where $I_c$ scales with the square-root of the sample
dimension~\cite{eastham_critcurr_2010}:
\begin{eqnarray}
  I_c &\sim& \frac{e \rho_s}{\hbar}
  \sqrt{\frac{L_x}{\ld}}\frac{L_y}{\ld}\quad
  (\mbox{quasi-1D:\ } L_x\ll \ld \ll L_y)\,.
  \label{eq:critcurrentq1d}\\
&\sim&
  \frac{e \rho_s}{\hbar} \sqrt{\frac{L_x L_y}{\ld^2}}
  \quad(\mbox{for\ } L_x, L_y \ll \ld).
  \label{eq:critcurrentq0d}
\end{eqnarray} 
This crossover provides an experimental probe of the domain size $\ld$.

We thank P.~B.~Littlewood for helpful discussions. This work was
supported by EPSRC-GB (EP/C546814/01) and Science Foundation
Ireland (SFI/09/SIRG/I1952).

\bibliographystyle{apsrev}
\bibliography{qhcritj}
\end{document}